# Breaking Community Boundary: Comparing Academic and Social Communication Preferences regarding Global Pandemics

Qingqing Zhou[1], Chengzhi Zhang[2, *]

1. Department of Network and New Media, Nanjing Normal University, Nanjing 210023, China
2. Department of Information Management, Nanjing University of Science and Technology, Nanjing 210094, China

**Abstract**
The global spread of COVID-19 has caused pandemics to be widely discussed. This is evident in the large number of scientific articles and the amount of user-generated content on social media. This paper aims to compare academic communication and social communication about the pandemic from the perspective of communication preference differences. It aims to provide information for the ongoing research on global pandemics, thereby eliminating knowledge barriers and information inequalities between the academic and the social communities. First, we collected the full text and the metadata of pandemic-related articles and Twitter data mentioning the articles. Second, we extracted and analyzed the topics and sentiment tendencies of the articles and related tweets. Finally, we conducted pandemic-related differential analysis on the academic community and the social community. We mined the resulting data to generate pandemic communication preferences (e.g., information needs, attitude tendencies) of researchers and the public, respectively. The research results from 50,338 articles and 927,266 corresponding tweets mentioning the articles revealed communication differences about global pandemics between the academic and the social communities regarding the consistency of research recognition and the preferences for particular research topics. The analysis of large-scale pandemic-related tweets also confirmed the communication preference differences between the two communities.

# 1 Introduction

Rapid changes in ecology, climate, population growth, and complex population mobility mean that pandemics will become more frequent and harder to prevent and contain (Bedford et al., 2019). Outbreaks of pandemics, such as SARS (Severe Acute Respiratory Syndrome) (He et al., 2004), MERS (Middle East Respiratory Syndrome) (Zumla et al., 2015), and COVID-19 (Corona Virus Disease 2019) (Kupferschmidt & Cohen, 2020), have caused immeasurable negative impacts on the global economy, politics, and society.

Due to the development of science and technology, researchers in many fields have responded quickly. They engaged in massive interdisciplinary cooperation from multiple research perspectives, resulting in a large number of research results. The research involves global health (Murdoch & French, 2020), antibody development (Tetro, 2020), and the impact of gender (Wenham et al., 2020) and age (Huang et al., 2020) on virus infection. High-quality research is widely disseminated in the academic community, enabling researchers to quickly understand the development process of pandemics and grasp information about the research frontier. However, in addition to the researchers, the general public is also in the midst of a global pandemic (i.e., COVID-19). People urgently need relevant information related to the pandemic to soothe their panic and to avoid psychological diseases, like depression or anxiety. Therefore, two-way information communication between academic and social communities is crucial. Do differences in information communication preferences between the two communities hinder their communication? Are there differences in communication preferences between academia and the public regarding pandemics and the related research results? In other words, can research results widely disseminated in the academic community cause large-scale discussion in the social community? Have the popular academic research topics attracted public attention or been recognized by the public?

To answer the above questions, we first collected scientific research related to pandemics, and we crawled social media to assess the discussion and dissemination of the pandemic and pandemic-related research. Then, the scientific research and corresponding user-generated content were mined using topic modeling and sentiment analysis to calculate the impact of the research on the academic and social communities, respectively. Meanwhile, concern about and recognition of different research topics among academia and the public were identified, and differences between research topics and social topics were mined. Finally, differences were detected in communication preferences about pandemics between the academic community and the social community.

## 2 Related work

Pandemics, especially the COVID-19 pandemic, have attracted the attention of global academic and social communities (Yin et al., 2021). Mass research groups, funding agencies, and volunteers are using their expertise and resources to support the fight against the current pandemic (Colavizza et al., 2020). Therefore, a vast amount of pandemic-related research has been published and then discussed and disseminated on multiple social web platforms. In this section, we describe two related categories: analysis of scientific research on pandemics based on academic studies and analysis of scientific research based on social media

2.1 Analysis of scientific research on pandemics

2.1.1 Scientific research on pandemics

Many researchers focused on the characteristic mining of the pandemic. Lu et al. (2020) analyzed the genomic characteristics and epidemiology of COVID-19. Deng et al. (2020) detected the clinical characteristics of patients who succumbed to and of patients who recovered from COVID-19. Zhao et al. (2020) compared the clinical features of COVID-19 pneumonia with other types of pneumonia and found that liver function damage was more frequent in COVID-19 than in non-COVID-19 patients. Chakraborty and Ghosh (2020) focused on generating real-time forecasts of future COVID-19 cases for multiple countries and identifying the demographic and disease characteristics of the pandemic. Chen and Yu (2020) used a second derivative model to characterize the COVID-19 pandemic and concluded that the pandemic appeared to be nonlinear and chaotic.

The impacts of gender and age on virus infection are also important research directions (Huang et al., 2020; Wenham et al., 2020). Zhou et al. (2020) reported increasing odds of in-hospital death associated with older age groups. Park et al. (2020b) found that patients over 70 were more likely to die from COVID-19 infection than younger individuals and that men had a higher case fatality rate than women. Infection and prevention in children have also been widely studied (Fang & Luo, 2020).

### 2.1.2 Scientific research on pandemics-related articles

The explosive growth in COVID-19 information presents challenges to researchers and the public in keeping up with emerging knowledge. Therefore, many researchers focused on the analytic induction of research articles in this domain. Park et al. (2020a) investigated obvious and less obvious consensus points on suitable drug treatments by analyzing 34 published articles on COVID-19 and drug-repurposing. Aristovnik et al. (2020) used innovative bibliometric approaches to conduct an extensive bibliometric analysis of COVID-19 research across the science and social science research landscape. Zhang et al. (2020a) characterized, quantified, and measured the response of academia to international public health emergencies by conducting a comparative bibliometric study of six infectious disease outbreaks since 2000, including COVID-19.

Many researchers focus on analyzing the topics of pandemic-related articles to detect current research trends and directions. Porter et al. (2020) reassembled COVID-19 topical content to address topical evolution research issues, aiming to make pandemic-related knowledge more accessible. Radanliev et al. (2020) conducted data mining of scientific literature records from the Web of Science Core Collection and presented visualizations of interrelationships between scientific research data records on COVID-19. Zhang et al. (2020b) analyzed topic evolution, disruption, and resilience in early COVID-19 research. They characterized terms featured in articles on early COVID-19 research and knowledge pathways using term extraction, evolutionary pathways, and statistical analysis. Also, many researchers are committed to conducting bibliometric analysis of pandemic-related articles to generate high-quality research results. Odone et al. (2020) conducted systematic screening, quantitative assessment, and critical appraisal of the first 10,000 scientific articles published on COVID-19 to identify the content, trends, and quality of scientific publishing. Elhawary et al. (2020) summarized the characteristics of the top 50 cited COVID-19-related publications that emerged early during the pandemic to determine the factors that lead to highly impactful publications. Torres-Salinas et al. (2020) compared the uptake and social media attention between open access COVID-19 related literature and subscription-based articles.

The above analysis indicates that academia has carried out a series of studies on the pandemic itself, such as those related to the treatment of the illness and the characteristics of the patients. Researchers also summarized research publications related to the pandemic to provide comparative and comprehensive pandemic knowledge, thereby helping researchers and the public to access and understand the relevant information more efficiently.

## 2.2 Analysis of scientific research based on social media

### 2.2.1 Scientific research based on social media

Digital era products like social web platforms (e.g., Twitter, Facebook, Wikipedia, and Weibo) generate a new type of research data that may reflect the more widespread value of scientific research (Metaxas & Mustafaraj, 2014; Rogers, 2013). Eysenbach (2011) used Twitter mentions to predict the number of citations and the scientific impact of papers over the first three days of their publication. Cosco (2015) measured research impact factors and forecast citations using Twitter tweets. Thelwall and Nevill (2018) proved that Mendeley reader counts could predict later citation counts for research papers. Thelwall et al. (2013), Ortega and Luis (2018), and Huang et al. (2018) found significant positive correlations between social media mentions (including tweets, Facebook wall posts, and blog mentions) and normalized research citations, enabling the evaluation of research impact. However, Bornmann (2015) found that the correlation between traditional citations and micro-blogging counts was negligible. Fausto and Aventurier (2016) conducted a bibliometric analysis of the scientific literature on Twitter. They also found no clear relation between the largest number of mentions of published articles on Twitter and a more significant number of citations received.

Meanwhile, social web data can also be used to analyze published research and to examine research ideas, methods, and ethics. Williams et al. (2013) identified and classified over 1,000 academic papers in Twitter data to conduct Twitter-based research. Bruns et al. (2014) conducted content analysis on 382 academic publications utilizing Twitter data to prompt reflexive evaluation of scientific research. Mohammadi and Thelwall (2014) suggested that Mendeley readership data could be used to help capture knowledge transfer across scientific disciplines. Lin and Zhang (2020) compared scholars' blog topics posted on academic social networking sites with the keywords of their published papers to help scholars make better use of academic social platforms for informal academic communication. Kolahi et al. (2020) analyzed scientific articles in the field of endodontology. They proved that research would have more influence if journals and researchers set up their own social media profiles to share research findings promptly and to communicate with their network and audience.

### 2.2.2 Pandemic research based on social media

Public attitudes toward pandemics are critical in reducing the spread of pandemics. Hence, it is important to assess people's reactions and the means of information dissemination, especially social media. Thelwall and Thelwall (2020) conducted a thematic analysis of frequently retweeted early COVID-19 tweets to investigate important issues shared on Twitter in the early stages of the public reaction to COVID-19. Cinelli et al. (2020) analyzed engagement and interest in COVID-19. They provided a differential assessment of the evolution of the discourse on a global scale on each social platform (including Twitter, Instagram, YouTube, Reddit, and Gab). Li et al. (2020) conducted content analysis on microblogs to detect disease distribution and public attitudes and behaviors. Gao et al. (2020) analyzed pandemic-related public opinion information on Weibo to identify the attitude of the public during different stages of the pandemic. Kabir and Madria (2020) used tweets as shared information to visualize topic modeling and human emotions during the COVID-19 pandemic. Ni et al. (2020) examined the impact of social media use on the mental health of community and health professionals. They concluded that caution is warranted regarding excessive time spent on COVID-19 news on social media. Chen et al. (2020c) collected user reviews of seven major online education platforms before and after the outbreak of COVID-19 to analyze the impact of the virus on user experience.

Meanwhile, social media provide an effective way to share information about the pandemic. Therefore, many researchers are committed to building and sharing data about the pandemic with social networks (Chen & Yu, 2020; Moorthy et al., 2020). Drew et al. (2020) developed a COVID-19 symptom tracker mobile application to enable rapidly scalable epidemiologic data collection and analysis. Vergoulis et al. (2020) constructed an open dataset about pandemic-related papers by providing various scientific impact indicators for the relevant papers. Chen et al. (2020a) described a multilingual coronavirus Twitter dataset that they have been continuously collecting since January 22, 2020. The dataset could help to track scientific coronavirus misinformation and unverified rumors.

The above social-web-based research indicates that social web platforms can provide abundant real-time information about real events (Yin et al., 2015). Social web users share and replicate such information freely and quickly in an intense interaction through several actions (e.g., read, forward, comment), revealing their communication preferences (Zappavigna, 2011). Meanwhile, there may be a difference in communication preference regarding scientific research between the academic community and the social community. Social media mentions do not always reflect the impact of highly cited research studies (Costas et al., 2015).

In short, researchers from different fields have carried out an enormous amount of interdisciplinary research on pandemics. What is the public's attitude toward these studies? Are scholars' research preoccupations consistent with the public's information needs? Currently, few existing studies analyze the differences between academic and social opinions or attitudes toward pandemics and related research results. An analysis of the differences between the two could help researchers understand the mental conditions and information needs of the public during the pandemic. This could help determine research directions and topics, leading to high-quality academic research with higher social value. Meanwhile, the public could access the latest high-quality research results in a timely manner, thereby expanding their knowledge reserves. Therefore, through the multidimensional analysis of pandemic-related articles, this paper compares the communication preference differences for scientific research between academic and social communities. The aim is to break the boundary between the two communities.

## 3 Research question

By comparing academic and social opinions and attitudes toward the global pandemic, this paper aims to answer the following questions to identify the communication differences regarding the pandemic between the two communities.

RQ1. Are academic and social communities consistent in their attitudes toward the recognition of scientific research? For example, are the scientific articles widely recognized by academia also of interest to the public?

RQ2. Are there any preference differences for pandemic-related topics? For example, are there any preference differences for scientific research on different research topics between the academic and social communities? Are the topics in scientific research similar to those in social media?

# 4 Methodology

## 4.1 Data

The corpus analyzed in this paper contains two main parts, namely, pandemic-related articles and Twitter data mentioning the articles. We collected the corpus from April to June 2020. First, we examined the CORD-19[1] open research dataset to access the pandemic-related articles.[2] Second, we enriched the CORD-19 dataset with data from Dimensions[3] (e.g., number of citations of the articles) and Altmetric[4] (e.g., number of mentions on social media) (Colavizza et al., 2020; Herzog et al., 2020). Finally, we collected Twitter data (e.g., Twitter content, tweeter information) based on Twitter IDs extracted from Altmetric data with Twitter API.[5] Table 1 shows an overview of the corpus used in this paper. We collected 51,843 articles related to global pandemics. Almost all the articles are listed in Dimensions, while about 30% of the articles are mentioned in Altmetric. About 17% of the articles are discussed on Twitter, generating about 927,266 tweets. In addition, we collected 58,937,258 tweets concerning the COVID-19 pandemic from January to April 2020 (not confined to Twitter data mentioning pandemic research) (Chen et al., 2020b).

**Table 1** Data statistics

| Source | Size |
| --- | --- |
| CORD-19 publications | 51,843 articles |
| Dimensions | 50,338 (97.10%) articles |
| Altmetric | 15,400 (29.71%) articles |
| Twitter | 9,088 (17.53%) articles, 927,266 tweets (mentioning the articles) |
| | 58,937,258 tweets about the pandemic |

## 4.2 Framework

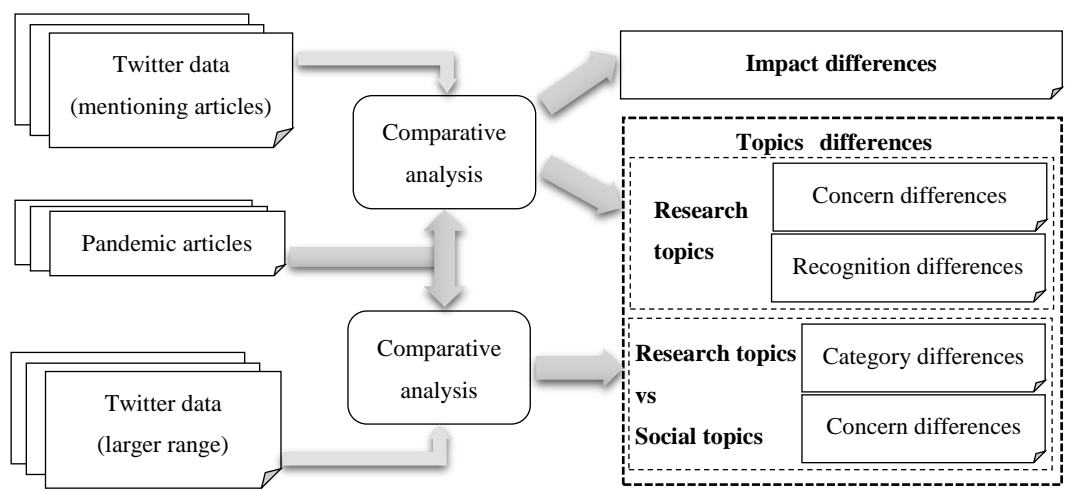

**Figure 1.** Analytic framework for communication preferences regarding pandemics

---

[1] https://pages.semanticscholar.org/coronavirus-research
[2] COVID-19 open research dataset (CORD-19). 2020. Version 2020-04-04. Semantic Scholar https://pages.semanticscholar.org/coronavirus-research. Accessed 2020-04-07.
[3] https://docs.dimensions.ai/dsl/
[4] https://api.altmetric.com/
[5] api.twitter.com

The primary purpose of this study is to detect communication differences between the academic community and the social community regarding global pandemics. The overall framework is shown in Figure 1. First, pandemic-related articles and corresponding Twitter data were collected. Second, the topics explored in the articles were extracted, and the sentiments of tweets mentioning these articles were identified. Meanwhile, we extracted topics explored in pandemic-related tweets (not confined to tweets mentioning pandemic research). Finally, comparative analysis was conducted, and comparison results from two dimensions (i.e., impact differences and topic differences) were obtained.

### 4.2.1 Topic extraction

Researchers may focus on multiple aspects of global pandemics and publish vast numbers of pandemic-related articles on different topics. In this paper, we used Latent Dirichlet Allocation (Hoffman et al., 2010) to extract the topics of these articles. We represented each article as a vector of topics. As an article may contain several topics, we took the topic with the highest probability as the final topic class of the article, as calculated in equation (1)

$$arti_{topic_i} = \max(p_{i1}, p_{i2}, \ldots, p_{in}) \tag{1}$$

where $arti_{topic_i}$ is the final topic class of article $i$, $p_{in}$ means the probability of article $i$ belonging to topic $n$. $n$ denotes the number of topics explored in pandemic-related articles. In this paper, perplexity was used to evaluate the performance of topic extraction based on different topic numbers to identify the optimal $n$ (Blei et al., 2003; Olivier & Helen, 2013). The topic extraction method used to analyze pandemic-related tweets is similar.

### 4.2.2 Sentiment classification

We conducted sentiment analysis on tweets mentioning pandemic-related articles to determine the public's attitudes toward these articles. Specifically, we collected tweets that mentioned pandemic-related articles. We then extracted sentiment indicators (i.e., positive and negative terms) based on a sentiment lexicon (i.e., SentiWordNet[6]). Finally, we calculated the sentiment scores of all tweets using equation (2) (Pang & Lee, 2008)

$$Senti_j = \frac{\#pos_j - \#neg_j}{\#pos_j + \#neg_j} \tag{2}$$

where $Senti_j$ is *sentiment score* of tweet $j$, $\#pos_j$ and $\#neg_j$ is the number of positive and negative terms in the tweet $j$, respectively.

### 4.2.3 Impact difference analysis

We calculated the impact of pandemic-related articles in academia and in social media, respectively, to measure the degree of consistency between academic and social attitudes to compare the communication differences between the two communities.

We computed the academic impact of articles using equation (3)

$$aca_{imp_i} = \begin{cases} \frac{\#citation_i}{t - t_i}, & t \neq t_i \\ \#citation_i, & t = t_i \end{cases} \tag{3}$$

where $aca_{imp_i}$ means *academic impact score* of article $i$, $\#citation_i$ denotes the citation number of article $i$. $t_i$ is the publication year of article $i$. $t$ means the current year, in this

---

[6] http://sentiwordnet.isti.cnr.it/

instance, 2020.

We computed the social impact of articles with two metrics, including *social sentiment score* and *social user score,* as calculated using equations (4) and (5)

$$soc\_senti_{imp_i} = \frac{1}{m_i}\sum_{j=1}^{m_i}(|Senti_{ij}| \times log(retweet_{ij})) \qquad (4)$$

where $soc\_senti_{imp_i}$ is *social sentiment score* of article $i$, $m_i$ denotes the number of tweets mentioning article $i$, $Senti_{ij}$ is *sentiment score* of tweet $j$, $retweet_{ij}$ is the number of retweets of tweet $j$.

$$soc\_user_{imp_i} = log(\sum_{j=1}^{n_i} follower_{ij}) \qquad (5)$$

where $soc\_user_{imp_i}$ is *social user score* of article $i$, $n_i$ denotes the number of unique users posting tweets mentioning article $i$, $follower_{ij}$ is the number of followers of user $j$.

4.2.4 Topic difference analysis

We analyzed the differences between the academic community and the social community regarding topic preferences from two dimensions: preference differences about research topics and differences about research topics and social topics. Research topics are topics extracted from pandemic-related articles, while social topics are topics extracted from pandemic-related tweets. Concern differences and recognition differences were analyzed for research topic preferences in the academic community and the social community. Concern differences were used to measure the difference in concern intensity between the academic and the social communities for different topics by comparing the number of articles and tweets on different topics. Specifically, we computed academic concerns about different topics using equation (6)

$$aca_{con_i} = \#articles_i/\#articles \qquad (6)$$

where $aca_{con_i}$ is *academic concern score* of topic $i$, $\#articles_i$ denotes the number of articles about topic $i$, and $\#articles$ means the number of academic articles.

We computed social concerns about different topics with three metrics in equations (7) to (9)

*Social concern metric 1*: $soc\_articles_{con_i} = \#articles\_tweet_i/\#articles_i \qquad (7)$

*Social concern metric 2*: $soc\_tweet_{con_i} = \#tweet_i/\#articles_i \qquad (8)$

*Social concern metric 3*: $soc\_user_{con_i} = \#user_i/\#articles_i \qquad (9)$

where $soc\_articles_{con_i}$ is *social concern score* of topic $i$ based on articles and $\#articles\_tweet_i$ denotes the number of articles discussed on Twitter in topic $i$. $soc\_tweet_{con_i}$ is *social concern score* of topic $i$ based on tweets, and $\#tweet_i$ denotes the number of tweets mentioning the articles about topic $i$. $soc\_user_{con_i}$ is *social concern score* of topic $i$ based on users and $\#user_i$ denotes the number of unique users discussing the articles in topic $i$.

Recognition difference refers to the difference in the degree of recognition between academic and social communities for different topics, thus reflecting their respective topic preference. Specifically, we compared the academic impact scores and the two social impact scores (i.e., *social sentiment score* and *social user score*) for articles about different topics.

Topic category differences and topic concern differences were analyzed to compare research topics and social topics. Topic category differences aimed to compare the topic perspectives disseminated by the academic community and the pandemic topic directions discussed by the social community to identify the consistency between academic research and social knowledge needs. Concern differences were used to measure the difference in concern intensity for different topics between the academic and the social communities. Specifically, we computed social concerns about different social topics using equation (10).

$$social_{con_i} = \#tweet_i/\#tweets \qquad (10)$$

where $social_{con_i}$ is *social concern score* of social topic $i$, $\#tweet_i$ denotes the number of tweets about social topic $i$, and $\#tweets$ means the number of pandemic-related tweets.

## 5 Results

5.1 Impact differences of research articles between the academic community and the social community

5.1.1 Performance evaluation of sentiment analysis

To evaluate the sentiment analysis performance of tweets mentioning pandemic-related articles, we annotated 1,000 tweets with three sentiment labels, namely positive, negative, and neutral. The annotated set includes 350 positive tweets, 300 negative tweets, and 350 neutral tweets. We then evaluated the classification results based on the annotated set. The evaluation results are shown in Table 2. Table 2 shows that the scores for all three evaluation indicators are higher than 0.88. We can conclude that the classification results are reliable. Finally, we had sentiment labels for all 927,266 tweets. Figure 2 presents the classification results. Neutral and positive tweets account for about 40%, while negative tweets are significantly fewer than the other two polarities, accounting for about 13%.

**Table 2** Performance of sentiment analysis

| Indicators | Macro Precision | Macro Recall | $F_1$ |
|---|---|---|---|
| Scores | 0.8924 | 0.8854 | 0.8889 |

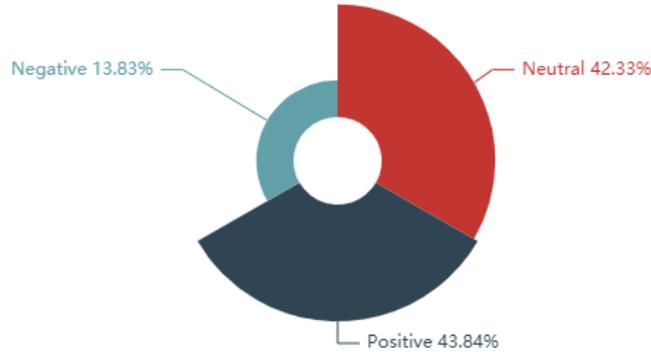

**Figure 2.** Sentiment classification results

5.1.2 Impact analysis on pandemic-related articles

Figure 3 shows the academic impact scores and the social impact scores for all 50,388 articles. Figure 3 shows that the academic impact scores for most articles are lower than 100. Social sentiment scores range from 0 to 3, and the scores for most articles are between 0 and 0.5. The social user scores for articles mentioned by social users are relatively evenly distributed around 4, and most articles have scores lower than 5. It can be concluded that there are similarities in the distributions of academic impact scores and social impact scores. The impact scores for most studies are low, and only a few studies get high scores. Meanwhile, the differentiated score distributions of the two social impact metrics indicate that analyzing the social impact of articles from multiple dimensions is necessary. The social sentiment scores represent users' attitudes to and attitude dissemination on the articles, measuring the social impact in depth. Meanwhile, the social user scores indicate the breadth of the social impact based on the number of users.

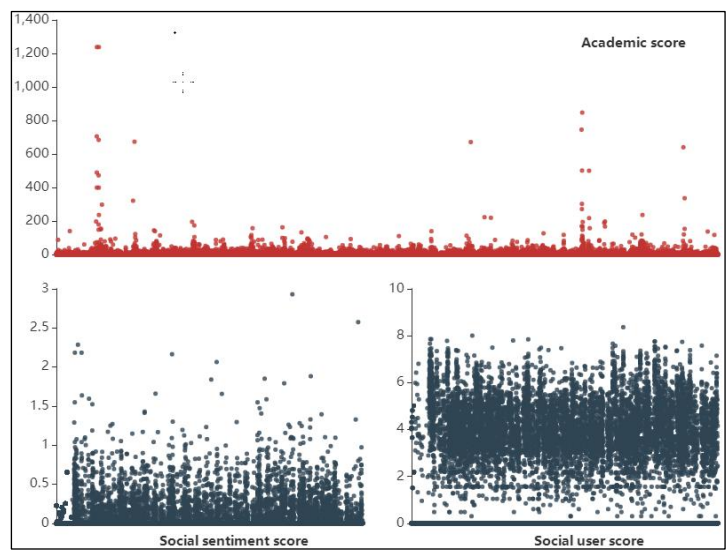

**Figure 3.** Statistics for the academic impact scores and the social impact scores

To detect the recognition consistency of scientific articles between academia and the public (i.e., to answer RQ1), we conducted a correlation analysis between the academic impact scores and the social impact scores. The results are shown in Table 3. Table 3 shows that the academic impact scores are significantly correlated with the social impact scores, while the correlation coefficients are low. It indicates that, with the exception of a small area of scientific research, the academic and social communities generally have inconsistent attitudes toward scientific articles. In other words, articles highly affirmed by academia may not be widely or actively disseminated in social media.

**Table 3** Correlations between academic impact scores and social impact scores

|  | Social sentiment scores | Social user scores |
| --- | --- | --- |
| Academic impact scores | 0.176*** | 0.312*** |

Notes: ***Significant at p = 0.001

This phenomenon reveals a mismatch between the research enthusiasm of the academic community and the knowledge needs of the social community, indicating the existence of a community boundary. We need to consider whether this mismatch is related to existing academic evaluation criteria. High-level academic research based on existing evaluation criteria may not meet the knowledge needs of the public. This phenomenon reminds us to take the current academic evaluation domain into consideration. What kind of academic research is good academic research? What kind of evaluation system is reasonable? Currently, most evaluation methods are based on academic resources, such as peer review and citations. Public recognition and communication have gradually become an important evaluation dimension. The comprehensive consideration of academic value and social value may become a standard for future evaluation of academic research. Meanwhile, media (especially social media) as a bridge connecting the academic and social communities need to play a better role in information communication.

## 5.2 Topic preferences of the academic community and the social community

To analyze the topic preferences of the academic and social communities (i.e., to answer RQ2), we mined topics from two dimensions: (1) comparison of the communication preferences for research topics in the two communities, and (2) comparison of communication differences between research topics and social topics. As shown in Figure 4, we compared the concern and recognition differences about topics explored in pandemic-related articles to identify the communication differences regarding academic research topics between two communities. Meanwhile, we collected more pandemic-related Twitter data (not confined to tweets mentioning pandemic articles) to detect a wider range of pandemic-related topics disseminated in the social community to compare the social topics and research topics from the aspects of topic category and concern.

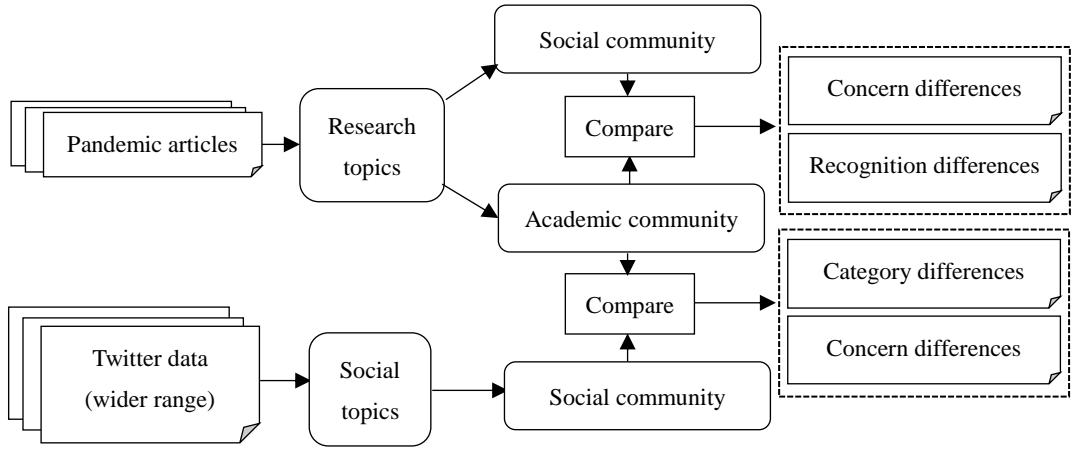

**Figure 4.** Topic preference comparison between the academic community and the social community

### 5.2.1 Communication preferences on pandemic-related research topics

**Table 4** Research topics of pandemic-related articles

| NO. | Topics | Keywords |
|---|---|---|
| 1 | Research on age differences in relation to the pandemic | Aged; Middle Aged; Animals; Adult; Child; Mice |
| 2 | Research on antibodies for the pandemic virus | Antibodies; Coronavirus Infections; Mice; Animals; Viral; Cell Line |
| 3 | Research on gender differences in relation to the pandemic | Female; Male; Humans; Viral; Coronavirus Infections; Mice |
| 4 | Global health research on the pandemic outbreak | Global Health; Disease Outbreaks; Humans; Animals; Viral; Mice |
| 5 | Research on the gene sequence of the pandemic virus | RNA; Molecular Sequence Data; Base Sequence; Coronavirus Infections; Disease Outbreaks; Cell Line |
| 6 | Research on the virus in the current pandemic | Viral; Coronavirus Infections; Pneumonia; Humans; Animals; Mice |
| 7 | Research on children in relation to the pandemic | Child; Infant; Humans; Viral; Animals; Mice |

In order to compare the preferences for different research topics in academic and social communities, we conducted topic extraction on 50,388 articles and compared extraction performance based on different numbers of topics (range from 1 to 10). Then, we had a list of seven topics. The topic extraction results are shown in Table 4. The seven topics include pandemic-related ages, antibodies, genders, global health, genes, viruses, and children. Based on the research articles on the seven topics, we can identify the directions and progress of academic research on the pandemic.

- *Concern differences of research topics*

To compare topic concern preferences between the academic community and the social community on academic research topics, we counted the distribution of the 50,388 articles among the topics to generate the academic concern scores, and we generated three social concern metrics, *social concern metric 1* (i.e., ratio of articles on the seven topics mentioned by tweets), *social concern metric 2* (i.e., numbers of tweets discussing the articles that were disseminated on Twitter), and *social concern metric 3* (i.e., numbers of users discussing the articles). The results are shown in Figure 5. A comparison of the academic concern scores and the scores for *social concern metric 1* reveals that the distribution trends of the seven topics are not consistent. In the academic concern area, Topic 1 (Research on age differences in relation to the pandemic) has the largest number of articles, followed by Topic 2 (Research on antibodies for the pandemic virus). Regarding the social concern metric 1, Topic 4 (Global health research on the pandemic outbreak) and Topic 7 (Research on children in relation to the pandemic), related articles are more likely to be mentioned on Twitter. Also, Topic 1 obtains the highest scores in *social concern metrics 2* and *3*, indicating that mass social media users have discussed Topic 1 in depth and have widely disseminated articles related to Topic 1 on social media (i.e., Twitter). Meanwhile, it is noteworthy that although the number of articles related to Topic 2 (Research on antibodies for the pandemic virus) is higher than for Topic 3 (Research on gender differences in relation to the pandemic) and Topic 4 (Global health research on the pandemic outbreak), the number of social media communications and the number of communicators present an opposite trend. The trend indicates that academia and the public have different preferences for the three topics.

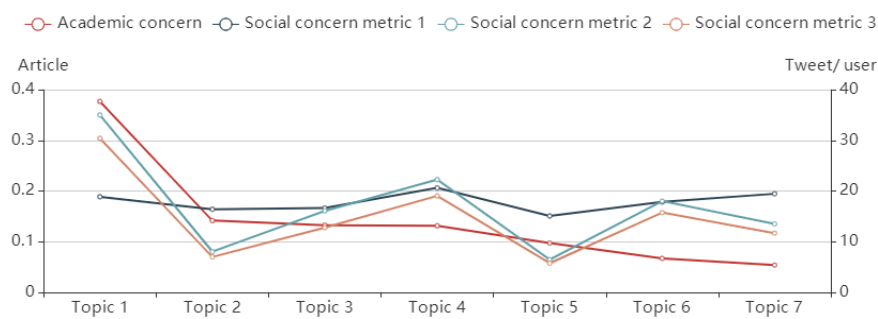

**Figure 5.** Academic concerns and social concerns about different topics

From the above analysis, it can be concluded that academia pays more attention to details and technology-related topics, while the public is more likely to be attracted by macro, general, and summative information. This situation may be related to the overall education level and social identity of the two communities. The academic community is generally highly educated. Also, the demands of professional and social responsibility mean that academics in biology, social science, and other fields can and must conduct detailed research on many aspects of the pandemic, thereby adding to knowledge of the phenomenon. However, due to the varying educational levels of the social community, some members of the public find detailed academic information difficult to understand. More importantly, they may not be interested in information on the technical aspects of the pandemic. They are more concerned with conclusive information (e.g., the current situation regarding the spread of the pandemic).

- *Recognition differences among research topics*

We hold that a high impact of an article implies high academic and social recognition. Therefore, to compare topic recognition preferences between the academic community and the social community, we compared the academic impact scores and the social impact scores of articles on seven topics using correlation analysis. The results are shown in Table 5.

Table 5 Correlations between academic impact scores and social impact scores for different topics

| Impact score | Topics | Social sentiment score | Social user score | N |
|---|---|---|---|---|
| Academic impact score | Topic 1 | 0.209*** | 0.312*** | 18,988 |
| | Topic 2 | 0.223*** | 0.361*** | 7,127 |
| | Topic 3 | 0.213*** | 0.349*** | 6,665 |
| | Topic 4 | 0.251*** | 0.361*** | 6,608 |
| | Topic 5 | 0.223*** | 0.347*** | 4,889 |
| | Topic 6 | 0.256*** | 0.367*** | 3,363 |
| | Topic 7 | 0.142* | 0.295*** | 2,698 |

Notes: ***Significant at p = 0.001, *Significant at p= 0.05

Table 5 shows that the correlation coefficients between the academic impact scores and the social impact scores of all seven pandemic-related research topics are low, indicating that there are differences in the degree of recognition of pandemic-related research between the academic and social communities. Except for Topic 7, the correlation coefficients and significance of the topics are similar. There is no obvious difference in the distribution of topic recognition for the six other topics between the academic community and the social community. In other words, while there are differences between the academic and social communities in topic recognition, the differences are universal and topic independent. The correlation coefficient and significance for Topic 7 are lower than those for the other topics. This indicates that the academic and social communities' recognition difference for research on Topic 7 is higher than for other research topics.

As previous analysis on topic concerns indicates, the academic and social communities have clear preference differences for pandemic research topics. Research topics that are highly praised in the academic community may not be widely disseminated by the social community. This may relate to the different information needs and evaluation systems of the different communities. In order to promote research on the pandemic, the academic community needs to engage in research on multiple aspects, then comment, evaluate, and optimize the research publications from an academic perspective. When massive research on different topics is published, only research results that are easy to understand and use will attract public comment and informed dissemination. On the one hand, this clear community boundary calls on the academic community to make research results more practical and suitable to meet real-world needs. On the other hand, in the Web 2.0 era, while active access by the public is a factor, the promotion and popularization of scientific knowledge is still the focus of academia, society, the educational community, and governments.

### 5.2.2 Communication preferences for pandemic-related research topics and social topics

To further compare pandemic-related communication preferences between the academic community and the social community on a larger data scale, we extracted the 1,000,000 tweets with the highest forwarding numbers every month from the 58,937,258 pandemic-related tweets. We analyzed the topics of 4,000,000 pandemic-related tweets. Table 6 shows the results of topic extraction. Five social topics were extracted: COVID-19 related infections, information, influences, workers, and virus.

Table 6 Topics of pandemic-related tweets

| NO. | Topics | Keywords |
|---|---|---|
| 1 | People infected with COVID-19 | Case; Confirm; First; Recover; Pneumonia; Coronavirus |
| 2 | COVID-19 related information | Coronavirus; Social distance; Lockdown rules; Emergency; Wuhan |
| 3 | The influence of COVID-19 on daily life | Social; Work; Rent; Family; Travel planning |
| 4 | COVID-19 related workers | Doctor; Nurse; Cleaner; Volunteer; WHO; Regulations |
| 5 | COVID-19 virus | Coronavirus; Wuhan; China; Symptom; Infect |

- *Category differences between research topics and social topics*

Table 4 showed the topics of pandemic-related research disseminated by the academic community. Table 6 showed the pandemic-related topics widely discussed by the social community. In Table 7, we classified the research topics and the social topics. The research topics can be classified into three categories: topics related to population characteristics, topics related to global health, and topics related to the nature of the virus. Social topics can also be classified into three categories: people, information, and the nature of the virus.

Table 7 Research topics and social topics

| Research topics | | Social topics | |
|---|---|---|---|
| Class 1: population characteristic | Topic 1: Research on age differences in relation to the pandemic | Class 1: people | Topic 1: People infected with COVID-19 |
| | Topic 3: Research on gender differences in relation to the pandemic | | Topic 4: COVID-19 related workers |
| | Topic 7: Research on children in relation to the pandemic | | |
| Class 2: global health | Topic 4: Global health research on the pandemic outbreak | Class 2: information | Topic 2: COVID-19 related information |
| | | | Topic 3: The influence of COVID-19 on daily life |
| Class 3: nature of the virus | Topic 2: Research on antibodies for the pandemic virus | Class 3: nature of the virus | Topic 5: COVID-19 virus |
| | Topic 5: Research on the gene sequence of the pandemic virus | | |
| | Topic 6: Research on the virus in the current pandemic | | |

The data in Table 7 indicate that in terms of topic categories, there are obvious differences between the topics of public concern and the topics of academic research. The topics discussed in the academic community involve in-depth research (e.g., population characteristics, gene sequencing, antibodies, etc.), while the topics discussed in the social community are more divergent, involving different groups of people and related information. Also, even in a similar topic class, such as the topic about the nature of the virus, the discussion perspectives of the two communities are different. The academic community tries to study coronavirus from the dimensions of virus infection, gene analysis, antibody research, and experiments on animals, while the social community discusses the conclusive information related to the pandemic from the dimensions of diagnosis and cure.

- *Concern differences between research topics and social topics*

We compared the topic concern preferences between the academic community and the social community on the global pandemic. We counted the concern scores on the pandemic-related research topics and social topics. The results are presented in Figure 6. Figure 6 shows that concern scores for the topic about infection, the topic about daily life, and the topic about the pandemic virus are each lower than the other. These topics are more and more in-depth and distanced from the social users' personal interests. However, in the academic community, in-depth topics are widely disseminated. The concern scores reveal that in the more extensive range of data related to the pandemic, the differential topic communication preferences of the academic and social communities are clear. The preference differences are consistent with the analysis results based on the pandemic-related articles. The social community pays more attention to self-interest-related and conclusive information than to micro and detailed information (e.g., the nature of the virus). In summary, there is a significant community boundary between the academic and social communities, whether it is based on the dissemination of scientific research or other information related to the pandemic.

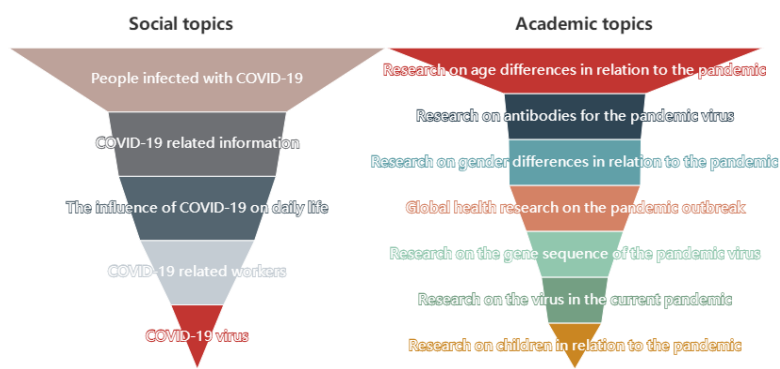

**Figure 6.** Topic concern scores of pandemic-related tweets and articles

## 6 Discussion

6.1 Comparative analysis of Altmetric scores with academic and social scores

The well-known bibliometric website, Altmetric.com, provides the bibliometric scores of academic publications by integrating the performance of publications (e.g., citations, downloads) on social platforms like Twitter and Google. Is there a significant difference between the Altmetric, academic, and social scores for articles? In other words, do the multidimensional scores calculated in this paper measure the value of these pandemic-related articles well? We conducted a correlation analysis of the three types of scores (i.e., Altmetric, academic, and social). The results are shown in Table 8. Table 8 shows that Altmetric scores are significantly correlated with academic scores and social scores. Articles with higher Altmetric scores tend to have higher academic values and social values, especially for recognition at the user level.

**Table 8** Correlations between Altmetric scores, academic scores, and social scores

| Scores | Academic impact score | Social scores | |
| --- | --- | --- | --- |
| | | Social sentiment score | Social user score |
| Altmetric scores | 0.438*** | 0.262*** | 0.604*** |

We compared social media mentions of pandemic-related articles provided by Altmetric.com and numbers of tweets and Twitter users discussing these articles. Table 9 shows the comparison results. We can see that Altmetric counts are highly correlated with tweet counts and user counts. Table 9 indicates that most discussions about the research articles on social platforms are posted on Twitter (in terms of the social media analyzed by Altmetric.com). In other words, the majority of users prefer to use Twitter to express their opinions of and attitudes to the pandemic. This finding also demonstrates the advantage of Twitter's instant reflection of high-interest current events and emergencies. Hence, we can conclude that social scores calculated by Twitter data can effectively measure the social performances of pandemic-related articles.

**Table 9** Correlations between Altmetric counts, tweet counts, and user counts

| Scores | #Tweets | #Twitter users |
|---|---|---|
| Altmetric counts | 0.999*** | 0.994*** |

## 6.2 Analysis of the community boundary

Communication preference usually reflects the level of knowledge and cognition about pandemics (Garner & Gillingham, 1991; Sotirovic, 2001). Members of the general public, represented by the majority of social media users, express concerns about whether "I (or people like me)" will be infected or how to effectively prevent pandemics. The focus of the general public is on results. However, the academic community pays more attention to the analysis of pandemics from the aspects of the nature of the virus, trying to analyze the causes and possible development trends of pandemics.

Therefore, we can state that the difference between the academic community and the social community, or the boundary between them, is the difference in their knowledge levels and social roles (Liao et al., 2014). The public does not have access to first-hand information. People can only receive pandemic-related information through the media. Therefore, most people experience problems identifying misinformation related to the pandemic. They believe and spread false information mostly out of fear of the pandemic, reflecting their low level of knowledge. The academic community has abundant data and resources and a high level of knowledge to devote to pandemic-related research. However, it seems that it is difficult for researchers to balance scientific research with the information needs of the public in a timely manner, thereby gradually creating a community boundary over the course of the pandemic. Therefore, in order to improve the social significance of scientific research, there is an urgent need to break the boundary between the two communities.

The development of science and technology has produced many communication channels, including social media platforms. These channels could enable the academic community to disseminate academic achievements more widely and promptly. This would not only help peers to understand the research frontier but also increase public access to knowledge and reduce the cost of knowledge acquisition. Meanwhile, the collective voice of the public could enable the academic community to determine the knowledge needs of the public, thereby setting research directions with higher social significance. Therefore, timely and effective open sharing of data (or information and knowledge) and the rational use of social network platforms (including information acquisition and screening, timely responses, and feedback) may be the key to eliminating the community boundary, thereby promoting the integration of the academic and social communities.

This study has some limitations. First, we collected metadata information on pandemic-related articles up to April 4, 2020, and we then collected relevant data (e.g., Altmetric data and Twitter data). However, the number of pandemic-related articles is increasing rapidly. Therefore, data expansion is needed in a follow-up study to obtain more comprehensive analysis results. Second, in the multidimensional analysis of the research articles, we did not distinguish the types of pandemics. Therefore, we cannot show a separate and detailed analysis of the COVID-19 pandemic. Hence, in the follow-up study, based on an expanded corpus, we need to distinguish between the types of pandemics to compare the characteristics of different pandemics and their corresponding research. Third, we did not determine the identity of social communicators (namely, Twitter users). Further research could identify the presence of researchers among the social communicators and establish the proportion of researchers to the general public. In the future, to further verify the conclusions of this paper, we intend to identify the categories of social communicators through the occupation information contained in the communicators' personal Twitter information. This paper conducted a preliminary analysis of the communication preferences of the academic and social communities. More state-of-the-art techniques and methods (topic extraction, sentiment classification, etc.) should be applied to produce finer-grained analysis results.

## 7 Conclusion

This study analyzed the attitudes and opinions of the academic and social communities on pandemic-related research to identify the communication differences between academia and the public on pandemics and the related research.

In answer to the first research question, the academic community and social community have inconsistent knowledge of pandemic-related research. Research that is highly recognized in academia may not be widely disseminated in the social community. Regarding the second research question, there are clear preference differences concerning research topics between the academic community and the social community. Academia pays more attention to concrete and detailed research related to pandemics (e.g., antibodies and the gender effect), while the public focuses more on the macro situation (e.g., global health). Meanwhile, in addition to the concern differences, recognition differences on research topics are evident, and the differences between research topics and social topics also reflect the existence of community communication preferences.

In conclusion, the methodology of this paper offers some suggestions for identifying communication differences between the academic and social communities on pandemic-related research and on breaking the community boundary. If research results are to have higher social value, researchers should conduct scientific research according to the information needs of the public. Meanwhile, the public can acquire and filter pandemic-related knowledge according to the academic evaluation results, thereby obtaining reliable information in a more efficient manner.

## Acknowledgments

This work is supported by Major Projects of National Social Science Fund (No. 16ZDA224), the National Social Science Fund Project (No. 19CTQ031) and the Nanjing Normal University Project (No.184080H202B208).